\documentclass[12pt]{iopart}

\usepackage{amsthm}
\usepackage{amssymb} 
\usepackage{mathrsfs}
\usepackage{multirow}
\usepackage{color}
\usepackage{setspace} 
\usepackage{graphicx} 
\usepackage{subfig} 
\usepackage{url} 
\usepackage{hyperref} 

\let\l=\left
\let\r=\right
\def\be{\begin{equation}}
\def\ee{\end{equation}}
\def\bea{\begin{eqnarray}}
\def\eea{\end{eqnarray}}

\begin{document}

\title[Galileon]{Fully consistent rotating black holes in the cubic Galileon theory}
\author{P.~Grandcl\'ement$^{1}$}
\vspace{10pt}
\begin{indented}
\item[$^{1}$] Laboratoire Univers et Théories, Observatoire de Paris, Université PSL, Université Paris Cité, UMR 8102 CNRS, F-92190 Meudon, France
\end{indented}
\ead{philippe.grandclement@obspm.fr}

\begin{abstract}

Configurations of rotating black holes in the cubic Galileon theory are computed by means of spectral methods. The equations are written in the 3+1 formalism and the coordinates are based on the maximal slicing condition and the spatial harmonic gauge. The black holes are described as apparent horizons in equilibrium. It enables the first fully consistent computation of rotating black holes in this theory. Several quantities are extracted from the solutions. In particular, the vanishing of the mass is confirmed. A link is made between that and the fact that the solutions do not obey the zeroth-law of black hole thermodynamics.

\end{abstract}


\section{Introduction}

In the last decade, observations have provided strong evidences that black holes are true astronomical objects. One cannot help mentioning the first observations of the gravitational waves resulting from the coalescence of two black holes in 2015 \cite{Abbott:2016} by the LIGO-Virgo collaboration \cite{LIGO, Virgo}. Since that first breakthrough, several tens of such events have been detected. Another strong evidence comes from the high angular resolution observations of the environment of the supermassive objects located at the center of galaxies. Those observations are made either in radio by the EHT collaboration \cite{EHT:2019} or in the infrared with the Gravity instrument \cite{GRAVITY:2020}.

If all the current observations are consistent with the compact objects being black holes described by General Relativity, it is believed that the next generation of gravitational wave detectors, LISA \cite{LISA, Amaro-Seoane:2022} or the Einstein Telescope \cite{Maggiore:2019}, will put more stringent constraints on their nature. Deviations from General Relativity could then be detected where black holes differ from the Kerr solution \cite{Kerr:1963}. In order to prepare the future observations and maximize the scientific payback, many research projects aim at computing black hole solutions in various alternative theories of gravity.

In a previous work \cite{VanAelst:2019}, black holes in the cubic Galileon theory were constructed. This theory of gravity belongs to the class of scalar-tensor theories known as Horndeski theories \cite{Horndeski:1974}, theories which lead to second-order field equations. It has been shown that black holes constructed in the cubic Galileon theory could differ from those obtained in General Relativity \cite{Babichev:2016}. Using quasi-isotropic coordinates, solutions of rotation black holes in this context were first obtained in \cite{VanAelst:2019}. However it turned out that the choice of quasi-isotropic coordinates was inconsistent because of a violation of the circularity condition (see Sec. 3.5 of \cite{VanAelst:2019} for a detailed discussion). It followed that the rotating solutions obtained were only approximate.

In order to cure the limitations of \cite{VanAelst:2019} one must move away from the quasi-isotropic coordinates. So, in this paper, one relies on numerical coordinates based on the maximal slicing condition and the spatial harmonic gauge. This choice has proven to enable the computation of various models of black holes, once appropriate boundary conditions are enforced on the fields at the horizon \cite{Grandclement:2022}. Here, the formalism of \cite{Grandclement:2022} is applied to the rotating black holes in the cubic Galileon theory and leads, for the first time, to exact (up to the numerical precision) solutions. 

The paper is organized as follows. Section \ref{s:equations} presents the theory and the various equations in the 3+1 framework. The formalism described in \cite{Grandclement:2022} is then briefly recalled. A detailed presentation of the resolution of the equation for the scalar-field is given, as it turned out to be the most difficult part in obtaining the solutions. Section \ref{s:results} presents various aspects of the computed black holes. After explaining the numerical setup, the achieved precision is assessed by showing the behavior of error indicators when increasing resolution. Last, some mathematical aspects of the configurations are discussed, in particular the fact that the black holes do not obey the zeroth-law of thermodynamics.

Throughout this paper Greek indices are four-dimensional ones, ranging from 0 to 3 whereas Latin indices are spatial ones, ranging from 1 to 3. Units such that $G=c=1$ are used.
 
 \section{Equations}\label{s:equations}
 \subsection{Model}
 
 Gravity in the cubic Galileon model involves a metric field ${\bf g}$ and a scalar-field $\phi$. The action contains an Einstein-Hilbert term, a kinetic one for the scalar-field and a non-standard contribution of higher order in $\phi$:

 \be
 \label{e:action_full}
 S \l({\bf g}, \phi\r) = \int \l[ \xi \l(R - 2\Lambda\r) - \eta \l(\partial \phi\r)^2 + \gamma \l(\partial\phi\r)^2 \Box \phi\r] \sqrt{-g} {\rm d}x^4.
 \ee

$g$ denotes the four-dimensional metric and $\nabla$ the covariant derivative associated to it. $\Lambda$ is the cosmological constant and $\xi$, $\eta$ and $\gamma$ some coupling constants. The kinetic term is $\l(\partial \phi\r)^2 = \nabla_\mu \phi \nabla^\mu \phi$ and $\Box \phi = \nabla_\mu \nabla^\mu \phi$.

Following \cite{VanAelst:2019}, the results presented in this paper are restricted to the case with $\Lambda=0$ and $\eta$ = 0. The action then only depend on one coupling constant $\gamma$ and reduces to 
 \be
 \label{e:action}
 S \l({\bf g}, \phi\r) = \int \l[R  + \gamma \l(\partial\phi\r)^2 \Box \phi\r] \sqrt{-g} {\rm d}x^4.
 \ee
The $\gamma$ appearing in Eq. (\ref{e:action}) relates to the one in Eq. (\ref{e:action_full}) by a simple rescaling by $\xi$ and the same notation is used for simplicity. The stress energy-tensor then reads
\be
\label{e:tmunu}
T_{\mu\nu} = \gamma\l[\partial_{\l(\mu\r.} \phi \partial_{\l.\nu\r)} \partial\phi^2 - \square\phi \partial_\mu \phi \partial_\nu \phi - \frac{1}{2}g_{\mu\nu} \partial^\rho\phi \partial_\rho \partial\phi^2\r].
\ee

The variation of the action with respect to the scalar-field leads to an equation that can be cast into the form of a current conservation $\nabla_\mu J^\mu=0$ with

\be
\label{e:current}
J_\mu = \gamma \l[\partial_\mu \phi \box \phi - \frac{1}{2} \partial_\mu \l(\partial\phi\r)^2\r].
\ee

Black holes that differ from the Kerr solution can be obtained by demanding that the scalar-field depends on time, in a linear manner 

\be
\label{e:ansatz}
\phi = q t + \Psi,
\ee
where $q$ is a constant and $\Psi$ is a time-independent field. The stress-energy tensor (\ref{e:tmunu}) contains only derivative of $\phi$ so that the form (\ref{e:ansatz}) leads to stationary solutions for which $q$ is a parameter. 

 \subsection{3+1 decomposition}\label{s:3p1}
 
 Throughout this work, a 3+1 decomposition of spacetime is used (see, for instance, \cite{Gourgoulhon:2007} for a detailed presentation of the formalism). Spacetime is foliated by spatial hypersurfaces of constant time $\Sigma_t$. On each slice, spatial coordinates $x^i$ are defined. In this context, the geometry of the full spacetime can be given in terms of several spatial quantities: a scalar the lapse $N$, a vector the shift $B^i$ and a three-dimensional metric $\gamma_{ij}$. The four-dimensional line-element then reads
 
\be
{\rm d}s^2 = (-N^2 + B_i B^i) {\rm d}t^2 + 2 B_i {\rm d}x^i {\rm d}t + \gamma_{ij} {\rm d}x^i {\rm d}x^j.
\ee

The normal to each hypersurface is given by $n_\mu = \l(-N, 0, 0, 0\r)$ and $n^\mu = \l(1/N, -B^i/N\r)$. All spatial indices (Latin) are manipulated by means of ${\bf \gamma}$. The second fundamental form, the extrinsic curvature tensor, is given by
\be
\label{e:extrinsic}
K_{ij} = \frac{1}{2N} \l(D_i B_j + D_j B_i - \partial_t \gamma_{ij}\r),
\ee
where $D$ denotes the covariant derivative with respect to $\gamma_{ij}$.

The various parts of the stress-energy tensor can be expressed in terms of $\Psi$ and the 3+1 quantities.
\be
\square \phi = \frac{1}{\sqrt{-g}} \partial_\mu \l( \sqrt{-g} T^\mu\r) = \frac{1}{N} D_i\l(NT^i\r),
\ee
with 

\be
T^i = g^{i\alpha} \partial_\alpha \phi = \frac{B^i}{N^2} q + \l(\gamma^{ij} - \frac{B^i B^j}{N^2}\r) D_j \Psi,
\ee
so that

\be
\label{e:boxphi}
\square \phi = \frac{1}{N} D_i \l[ \frac{B^i}{N} q + N  \l(\gamma^{ij} - \frac{B^i B^j}{N^2}\r) D_j \Psi \r].
\ee

One also has 
\be
\label{e:dphisquare}
\partial \phi^2 = g^{\mu\nu} \partial_\mu \phi \partial_\nu \phi \\
		= -\frac{q^2}{N^2} + 2 \frac{B^i}{N^2} q D_i \Psi + \l(\gamma^{ij} - \frac{B^i B^j}{N^2}\r) D_i \Psi D_j \Psi.
\ee

The last term appearing in Eq. (\ref{e:tmunu}) is

\be
\label{e:dphidphisquare}
\partial^\rho\phi \partial_\rho \partial\phi^2 = \frac{B^i}{N^2}q D_i \l(\partial \phi^2\r) + \l(\gamma^{ij} - \frac{B^i B^j}{N^2}\r) D_i \Psi D_j \l(\partial\phi^2\r).
\ee

The 3+1 projections of the stress-energy tensor can be written in terms of $\Psi$ also. The computation of $E = n^\mu n^\nu T_{\mu\nu}$ involves the following terms

\bea
n^\mu n^\nu \partial_{\l(\mu\r.} \phi \partial_{\l.\nu\r)} \l(\partial\phi^2\r) &=& -q \frac{B^i}{N^2} D_i \l(\partial\phi^2\r) + \frac{B^iB^j}{N^2} D_i \Psi D_j \l(\partial\phi^2\r)\\
n^\mu n^\nu \partial_\mu \phi \partial_\nu \phi &=& \l(\frac{q}{N} - \frac{B^i}{N} D_i \Psi\r)^2 \\
n^\mu n^\nu g_{\mu\nu} &=& -1.
\eea

It leads to 
\be
\label{e:E}
E = \gamma\l[-q \frac{B^i}{N^2} D_i \l(\partial\phi^2\r) + \frac{B^iB^j}{N^2} D_i \Psi D_j \l(\partial\phi^2\r) - \square\phi  \l(\frac{q}{N} - \frac{B^i}{N} D_i \Psi\r)^2
+ \frac{1}{2} \partial^\rho\phi \partial_\rho \partial\phi^2\r].
\ee

The computation of $P_i = -n^\mu \gamma^\nu_i T_{\mu\nu}$ involves the following terms

\bea
n^\mu \gamma^\nu_i  \partial_{\l(\mu\r.} \phi \partial_{\l.\nu\r)} \partial\phi^2 &=& \frac{q}{2N} D_i \l(\partial\phi^2\r) - \frac{B^j}{2N}\l(D_i\Psi D_j \l(\partial\phi^2\r) + D_j\Psi D_i \l(\partial\phi^2\r)\r) \\
n^\mu \gamma^\nu_i \partial_\mu \phi \partial_\nu \phi &=& \l(\frac{q}{N} - \frac{B^j}{N}D_j \Psi\r) D_i \Psi \\
n^\mu \gamma^\nu_i g_{\mu\nu} &=& 0.
\eea

One then finds 
\bea
\label{e:P}
P_i = - \gamma\l[\frac{q}{2N} D_i \l(\partial\phi^2\r) \r. &-& \frac{B^j}{2N}\l(D_i\Psi D_j \l(\partial\phi^2\r) + D_j\Psi D_i \l(\partial\phi^2\r)\r) \\
\nonumber &-& \l.  \square \phi \l(\frac{q}{N} - \frac{B^j}{N}D_j \Psi\r) D_i \Psi\r].
\eea

The last projection is $S_{ij} = \gamma^\mu_i \gamma^\nu_j T_{\mu\nu}$ with terms

\bea
\gamma^\mu_i \gamma^\nu_j \partial_{\l(\mu\r.} \phi \partial_{\l.\nu\r)} \partial\phi^2 &=& \frac{1}{2} \l(D_i\Psi D_j \l(\partial\phi^2\r) + D_j \Psi D_i \l(\partial\phi^2\r)\r) \\
\gamma^\mu_i \gamma^\nu_j \partial_\mu \phi \partial_\nu \phi &=& D_i \Psi D_j \Psi \\
\gamma^\mu_i \gamma^\nu_j g_{\mu\nu} &=& \gamma_{ij},
\eea

so that
\be
\label{e:S}
S_{ij} = \gamma\l[\frac{1}{2} \l(D_i\Psi D_j \l(\partial\phi^2\r) + D_j \Psi D_i \l(\partial\phi^2\r)\r) - \square \phi D_i \Psi D_j \Psi - \frac{1}{2} \gamma_{ij} \partial^\rho\phi \partial_\rho \partial\phi^2 \r].
\ee

The equation for the scalar-field can also be expressed in terms of the 3+1 quantities. The components of the conserved current (\ref{e:current}) read

\bea
J_0 &=& \gamma q \square\phi \\
J_i &=& \gamma \l[D_i \Psi \square \phi - \frac{1}{2} D_i \l(\partial\phi^2\r)\r], 
\eea
so that 

\be
\label{e:current_3p1}
J^i = g^{i\mu} J_\mu = \gamma\l[\frac{B^i}{N^2} q \square \phi + \l(\gamma^{ij} - \frac{B^i B^j}{N^2}\r) \l(D_j \Psi \square \phi - \frac{1}{2} D_j \l(\partial\phi^2\r)\r)\r].
\ee

Beware that ${\bf J}$ is not a three-dimensional vector so that $J^i$ doesn't simply relate to $J_i$ by a contraction with $\gamma_{ij}$. The conservation-law for the current is then

\be
\label{e:conservation}
D_i \l(N J^i\r) = 0.
\ee

 \subsection{Gravitational sector} \label{s:gravitational}
 
In \cite{VanAelst:2019} the field equations were solved by making use of quasi-isotropic coordinates. The unknowns of the numerical code were then the non-vanishing components of the various tensors (i.e. the $g_{rr}=g_{\theta\theta}$, $g_{\varphi\varphi}$ and $B^\varphi$ ones). This is to be contrasted with what is used here, which is an application of the method presented in \cite{Grandclement:2022}. The unknowns are the tensors $N$, $B^i$ and $\gamma_{ij}$ themselves and not their individual components. Maximal slicing and spatial harmonic gauge are used. Such choice of coordinates is enforced by modifying the original system of the 3+1 equations. Maximal slicing is a condition on the choice of time-coordinate. It amounts to maximizing the volume of the foliation hypersurfaces (see Sec. 9.2.2 of \cite{Gourgoulhon:2007}). From the mathematical point of view it translates in the fact that the trace of the extrinsic curvature tensor vanishes : $\gamma^{ij} K_{ij} \equiv K = 0$. This condition is enforced by removing all the occurrences of $K$ in the equations. 

The spatial harmonic gauge defines the choice of spatial coordinates. It translates in the condition that $V^i \equiv \gamma^{kl} \l(\Gamma_{kl}^i - \bar{\Gamma}_{kl}^i\r) = 0$, where $\Gamma_{kl}^i$ denotes the Christoffel symbols of $\gamma_{ij}$ and $\bar{\Gamma}_{kl}^i$ the ones of a background metric. In this paper the background metric is chosen to be the flat one $f_{ij}$ (see Sec. II.A of \cite{Grandclement:2022} for more details). In the equations, the occurrences of the Ricci tensor are replaced by $R_{ij} - \frac{1}{2} \l(D_i V_j + D_j V_i\r)$. This modification ensures that the second order derivatives of the metric appear as a Laplacian-like operator $\gamma^{kl} \partial_k \partial_l \gamma_{ij}$. The resulting system of equations is 

\bea
\label{e:hamilton_ok}
R - D_k V^k - K_{ij} K^{ij} &=& 16 \pi E \\
\label{e:momentum_ok}
D^j K_{ij} &=& 8 \pi P_i \\
\label{e:evol_ok}
\mathcal{L}_{\vec{B}} K_{ij} - D_i D_j N + &N& \l(R_{ij} - \frac{1}{2} \l(D_i V_j + D_j V_i\r) - 2 K_{ik} K_j^{\phantom{j}k} \r) \\
\nonumber
&=& 4 \pi  N  \l( 2 S_{ij} - \l(\gamma^{kl} S_{kl} - E\r) \gamma_{ij}\r) ,
\eea
where $K_{ij} = \frac{1}{2N} \l(D_i B_j + D_j B_i\r)$ (no time derivative in that case) and $\mathcal{L}$ denotes the Lie derivative. Equation (\ref{e:hamilton_ok}) is the Hamiltonian constraint, Eq. (\ref{e:momentum_ok}) the momentum one and Eq. (\ref{e:evol_ok}) the evolution equation for $K_{ij}$. The modified system Eqs. (\ref{e:hamilton_ok}-\ref{e:evol_ok}) is expected to be well posed. However, for the solution to be a true solution of Einstein's equations, one must check, a posteriori, that the quantities $K$ and $V^i$ are indeed zero. This criterion has proven, in the past, to be a very strong test to assess the validity of solutions (see the tests performed in \cite{Grandclement:2022}).
 
 The presence of the black hole is imposed by enforcing appropriate boundary conditions on an inner sphere of radius $r_H$ and by solving Eqs. (\ref{e:hamilton_ok}-\ref{e:evol_ok}) outside that sphere. The boundary conditions used are those proposed in \cite{Grandclement:2022}. Only the main features of those conditions are recalled here and the reader should refer to \cite{Grandclement:2022} for a detailed presentation of the method. The boundary conditions encode the fact that the inner sphere is an apparent horizon, in equilibrium, and rotating at a velocity $\Omega$, a parameter that enters the condition of the shift. Some quantities are freely specifiable on the horizon, due to the fact that the coordinates used are defined by differential conditions (i.e. the equations $K=0$ and $V^i=0$ lead to differential equations on the fields). In this work, those quantities are chosen, at the inner boundary, as follows: $N=0.5$ and $\gamma_{r\theta} = \gamma_{\varphi\varphi}=0$. The spherically symmetric part of $\gamma_{rr}$ (i.e. its $Y_0^0$ component) is also free and chosen to be $8$. Those values have proven in the past (see \cite{Grandclement:2022}) to facilitate the convergence of the numerical code. Changing them would only lead to different choices of coordinates and not to new configurations.
 
The boundary condition on the shift radial component is $B^r = N \tilde{s}^r$, where $\tilde{s}^i$ denotes the unit normal to the horizon (with respect to the metric $\gamma_{ij}$). This condition comes from the fact the horizon is not expanding. It has an important implication because it makes some components of Eq. (\ref{e:evol_ok}) degenerate. This means that the factor in front of the highest order radial derivatives term $\partial_r^2 \gamma_{ij}$ vanishes on the horizon, implying that the equation becomes first order there and so does not require any boundary conditions to be solved. This in particular the case for the components $\l(\theta\theta\r)$, $\l(\theta,\varphi\r)$ and $\l(\varphi,\varphi\r)$. This point is also relevant for the scalar-field equation as discussed in Sec. \ref{s:field}. Once again, a detailed analysis can be found in \cite{Grandclement:2022}.
 
 Outer boundary conditions are given at spatial infinity and simply amount to demanding that the spacetime is asymptotically flat. This leads to $N=1$, $B^i=0$ and $g_{ij} = f_{ij}$, where $f_{ij}$ is the flat three-dimensional metric. From the numerical point of view, the use of a compactification in $1/r$ allows for those conditions to be enforced at exact spatial infinity (see Sec. \ref{ss:numerics}).
 
 \subsection{Field equation}\label{s:field}
 
The equations presented in Sec. \ref{s:3p1} involve only derivatives of the field $\Psi$. This is true for the various projections of the stress-energy tensor and also for the scalar-field equation (\ref{e:conservation}). In all those cases, the only relevant quantity is $D_i \Psi$. For axially symmetric solutions (the ones considered here) one has $D_\varphi \Psi=0$. So, from the numerical point of view, one can consider the two quantities $\Psi_r = \partial_r \Psi$ and $\Psi_\theta = \frac{1}{r}\partial_\theta \Psi$ as being the unknowns of the problem (a similar technique is used in \cite{VanAelst:2019}). The factor $1/r$ appearing in $\Psi_\theta$ comes from the fact that an orthonormal spherical tensorial base is used ; it ensures that $D_i \Psi = \l(\Psi_r, \Psi_\theta, 0\r)$.
 
Let us first consider the non-rotating case for which the solution is spherically symmetric. It follows that $\Psi_\theta=0$ and only one unknown  remains: $\Psi_r$. One needs to assess the order of Eq. (\ref{e:conservation}) in terms of the derivatives of $\Psi_r$. Equation (\ref{e:boxphi}) contains first order derivatives of $\Psi_r$. However, the prefactor in front of $D_r \Psi_r$ is $\l(g^{rr} - \frac{B^rB^r}{N^2}\r)$, which vanishes on the horizon, due to the boundary conditions used (see Sec. \ref{s:gravitational} and \cite{Grandclement:2022}). So $\Box \phi$ is a first order equation in terms of $\Psi_r$ but only zeroth order near the horizon. The same is true for Eq. (\ref{e:dphisquare}). It is then easy to verify that the conservation equation (\ref{e:conservation}) contains second order derivatives of $\Psi_r$ but is degenerate on the horizon, where only first order derivatives are present. It follows that only one boundary condition must be prescribed when solving Eq. (\ref{e:conservation}). A naive choice would be to relax any condition on the horizon and simply demand that $\Psi_r=0$ at infinity. However this leads to solutions that do not verify $K=0$ and $V^i=0$ and so are not solutions of Einstein's equations. A suitable choice consists in relaxing the condition at infinity and to enforce, on the horizon, the condition $V^r=0$. It is not trivial that this condition is sufficient to ensure that $K$ and $V^i$ vanish everywhere, but it turns out to be the case. Moreover the obtained solutions do indeed fulfill $\Psi_r=0$ at infinity. This situation concerning the boundary conditions is to be contrasted with what happens for the hairy black holes constructed in Sec. V of \cite{Grandclement:2022} where only a vanishing boundary condition at infinity is needed. One can conjecture that this difference comes from the fact that the equations are of different order, in terms of the scalar-field: second order for the hairy black holes of \cite{Grandclement:2022} and third order in the cubic Galileon case.

The above conclusions still hold in the rotating case, where both $\Psi_r$ and $\Psi_\theta$ must be taken into account. The conservation equation (\ref{e:conservation}) is solved using a single boundary condition on the horizon: $V^r=0$. An additional equation is provided by the symmetry of second derivatives: $\partial_r \l(r \Psi_\theta \r)= \partial_\theta \Psi_r$. From a technical point of view, this can be seen as a first order differential equation on $\Psi_\theta$, which is solved by demanding that, at infinity $\Psi_\theta=0$. It turns out that this is sufficient to lead to valid solutions. In particular, there is no need to impose anything for $V^\theta$ on the horizon. All this discussion about the scalar-field equation may seem rather technical but it is be noted that solving it properly has been the main difficulty in getting the solutions presented here.
 
 \section{Results}\label{s:results}
 \subsection{Numerical method}\label{ss:numerics}
 
 The equations presented in Sec. \ref{s:equations} are solved by means of the Kadath library \cite{Kadath, Grandclement:2009}. The setting is very similar to the one used in Sec. V of \cite{Grandclement:2022}. Space is divided into several spherical shells, the last one extending up to infinity thanks to a compactification in $1/r$. This setting is very standard and has proven to lead to good numerical results, in terms of both convergence and precision, in many different physical situations. Spherical coordinates $\l(r,\theta,\varphi\r)$ are used for the points and the tensors are given on the associated spherical orthonormal tensorial base. In each domain, a spectral decomposition of the fields is used where the angles $\l(\theta,\varphi\r)$ are expanded onto trigonometrical functions and the radial coordinate is described by means of Chebyshev polynomials. The numbers of points in each dimension are labelled $N_r$, $N_\theta$ and $N_\varphi$. This work is concerned with axisymmetric configurations only, so that the resolution in $\varphi$ is maintained fixed to $N_\varphi = 4$ (for technicalities in the Kadath library it is not possible to have less points).

The Kadath library enables to transform the system of equations into a discretized system on the spectral coefficients by means of a weighted residual method, essentially a version of the tau-method. Some regularities (i.e. on the axis of the spherical coordinates) are enforced by means of Galerkin basis. The non-linear discretized system is solved by means of a Newton-Raphson iteration. The code is parallelized using MPI and an typical job runs on 200 cores.

As in \cite{VanAelst:2019}, as a first step, the test-field solution is obtained. It corresponds to the limit $\gamma \rightarrow 0$, where the back-reaction of the scalar-field onto the metric sector is neglected. Thus the metric fields are fixed to the ones of a Schwarzschild black hole, obtained numerically in the maximal slicing and spatial harmonic gauge. It corresponds to the configurations obtained in Sec. III of \cite{Grandclement:2022}, with $\Omega=0$. Once the metric fields are known, the scalar-field is obtained by solving the equation $J_r=0$, which, in the non-rotating case, is equivalent to solving Eq. (\ref{e:conservation}). The test-field solution is known to have a different asymptotic behavior, at spatial infinity, than the full solutions. Indeed, as can be seen, for instance, on Eq. (41) of \cite{VanAelst:2019}, $\Psi_r$ behaves like $1/\sqrt{r}$ when $r\rightarrow \infty$. This square-root behavior is inconsistent with the compactification used by the Kadath library so that $J_r=0$ is solved only up to a finite radius $r_{\rm out}$. At that outer radius, the boundary condition $\partial_r \Psi_r + 2/r \Psi_r=0$ is enforced. This ensures that the true test-field solution is recovered when $r_{\rm out} \rightarrow \infty$. This technicality is only used for the test-field case and standard compactification and exact boundary conditions at infinity are used otherwise.

The test-field solution is used as a first initial guess to get solutions of the full Einstein-Klein-Gordon system. In the context of this paper, the coordinate radius of the black hole is maintained fixed to $r_H =1$. The parameter $q$ is also fixed to $q=1$, so the configurations depend on two remaining parameters: the angular velocity $\Omega$ and the coupling constant $\gamma$. Sequences are constructed by slowly varying those parameters. Let us mention that the various quantities are not scaled in the same way as in \cite{VanAelst:2019}. Indeed, in that previous paper, the various quantities were scaled by means of the radius of the horizon, which is not a coordinate independent quantity. In this paper, the circumferential radius $r_{\rm circ}$ of the black hole, that is the proper length of the horizon in the orbital plane divided by $2 \pi$, is used. This change in scaling makes a precise comparison between the two papers somewhat difficult. However this drawback is overcome by  the advantage in dealing with coordinate independent quantities only. Let us mention that the computation of $r_{\rm circ}$
involves the value of $\gamma_{\varphi\varphi}$ on the horizon, so that it is not constant for all configurations, even if $r_H$ is fixed.

 \subsection{Exact solutions with rotation}
 
\begin{figure}
    \begin{center}
        {
            \includegraphics[width=0.47\textwidth]{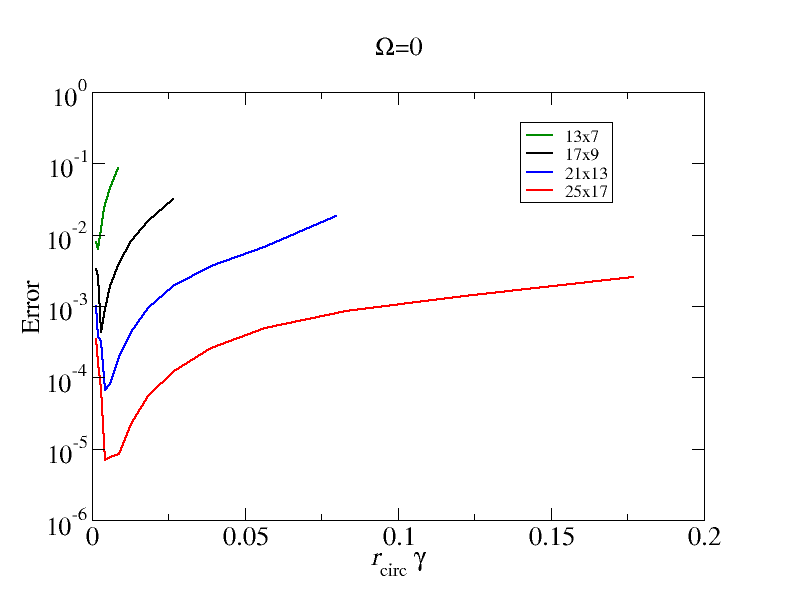}
        }
        {
            \includegraphics[width=0.47\textwidth]{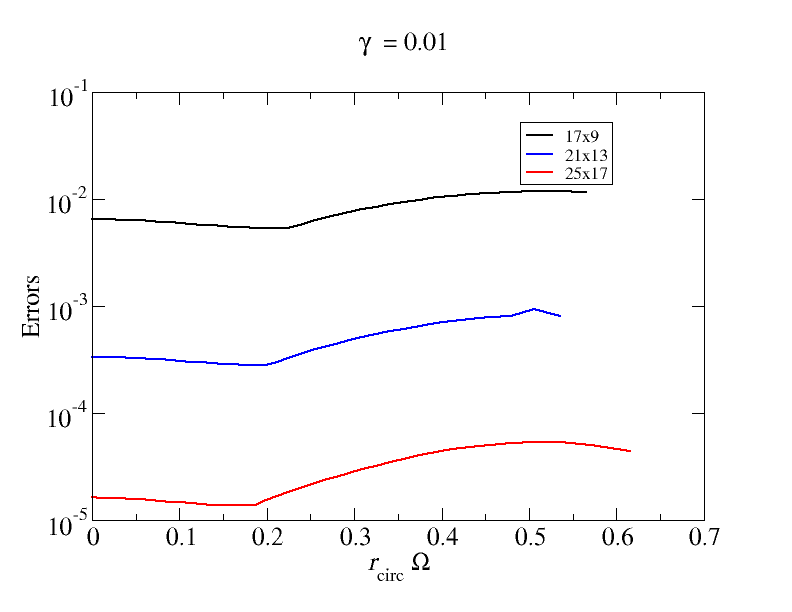}
        }
    \end{center}
\caption{Errors on the full set of equations, for various resolutions. The first panel shows the errors for the non-rotating solutions, as a function of the coupling constant $\gamma$ and the second one the errors for rotating configurations, as a function of the angular velocity $\Omega$. Spectral convergence is observed in both cases.}
\label{f:errors}
\end{figure}

 In order to assess the validity of the computed configurations, the various sources of numerical errors are monitored, as a function of the resolution. More precisely they consist of every equation that must be verified for a true solution but that is not solved explicitly by the code. This is the case for the gauge conditions $K=0$ and $V^i=0$ and for the $Y_0^0$ part of the expansion condition on the horizon $\Theta=0$  (see \cite{Grandclement:2022} and Sec. \ref{s:gravitational} for more details). An additional source of error comes from the method used to solve the partial differential equations (i.e. the tau-method). In order to enforce appropriate boundary and matching conditions, the last coefficients of the residual of the equations are not forced to be zero. Nevertheless, their value must go to zero with increasing resolution for well-posed problems. The maximal value of all those errors are shown in Fig. \ref{f:errors}, for various resolutions. The curves are labelled by the number of points in $r$ and $\theta$ in the form $N_r \times N_\theta$. The number of points in the radial direction is always higher than the angular one, a typical feature when using the Kadath library, that is known to facilitate convergence.
 
The first panel of Fig. \ref{f:errors} shows the errors for non-rotating configurations, as a function of the coupling constant $\gamma$. The errors exhibit a spectral convergence with an order of magnitude improvement between the various resolutions. Moreover, the code can reach higher values of the coupling constant with higher resolution. The second panel of Fig. \ref{f:errors} shows the errors for configurations with rotation. The coupling constant is maintained fixed to a moderate but not negligible value. The errors are plotted as a function of the angular velocity $\Omega$. As for the non-rotating case, the curves exhibit a clear spectral convergence. This is to be contrasted with the second panel of Fig. 1 of \cite{VanAelst:2019}, where the errors were independent of the resolution. This shows that the coordinates used in this work are consistent with the true rotating solutions, contrary to the quasi-isotropic ones (see discussion in Sec. 3.5 of \cite{VanAelst:2019}). The configurations presented in this work are the first exact numerical solutions of rotating black holes in the cubic Galileon theory.

 \begin{figure}
    \begin{center}
        {
            \includegraphics[width=0.3\textwidth]{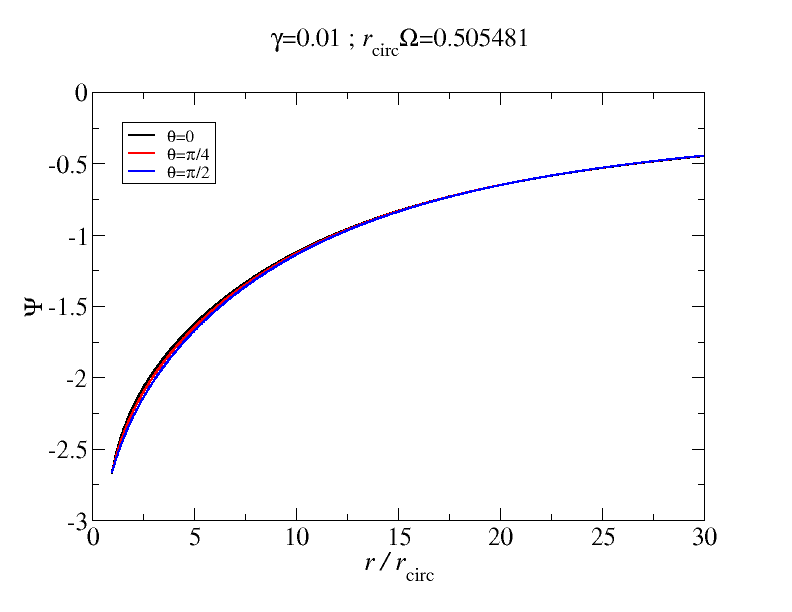}
        }
        {
            \includegraphics[width=0.3\textwidth]{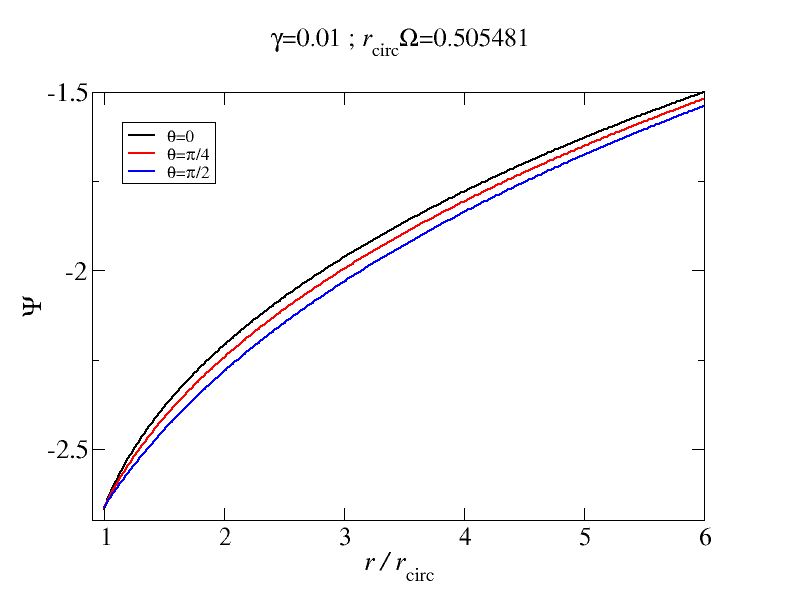}
        }
           {
            \includegraphics[width=0.3\textwidth]{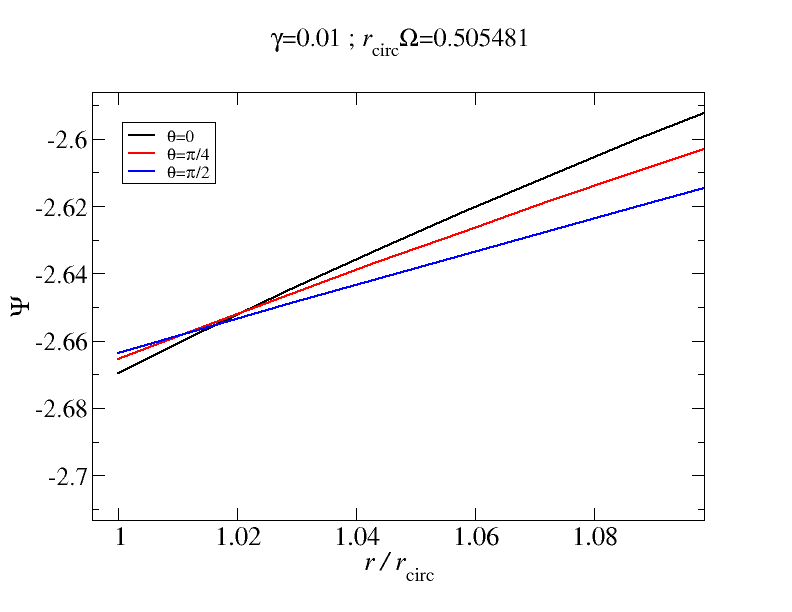}
        }
    \end{center}
\caption{Radial profiles of the scalar-field $\Psi$, for $\gamma=0.01$ and $r_{\rm circ}\Omega \approx=0.505481$. Three different values of the coordinate $\theta$ are shown in each case. The curves are shown as a function of $r / r_{\rm circ}$ and the three panels represent different radial regions.}
\label{f:prof_phi}
\end{figure}

The scalar-field $\Psi$ can be constructed from $\Psi_r$ by a numerical integration of the equation $\partial_r \Psi = \Psi_r$. This is easily done using Kadath. One can check that the solution is, as expected, regular everywhere. In particular no divergences appear on the horizon and the field vanishes at spatial infinity. As an illustration, some radial profiles of $\Psi$ are shown in Fig. \ref{f:prof_phi}. One can notice that the angular dependence is relatively small. This is to be expected as the amplitude of $\Psi_\theta$ is, in that case, almost two orders of magnitude smaller that the one of $\Psi_r$. Nevertheless some effect of the coordinate $\theta$ can be seen, in particular in the region close to $r \approx 3 r_{\rm circ}$ where the amplitude of $\Psi_\theta$ is the biggest. The value of the field on the horizon also exhibits some $\theta$ dependence. By carefully exploring the parameter space, it should be possible to find configurations for which the effect of $\theta$ is bigger but this is beyond the scope of this paper.

 \begin{figure}
    \begin{center}
        {
            \includegraphics[width=0.47\textwidth]{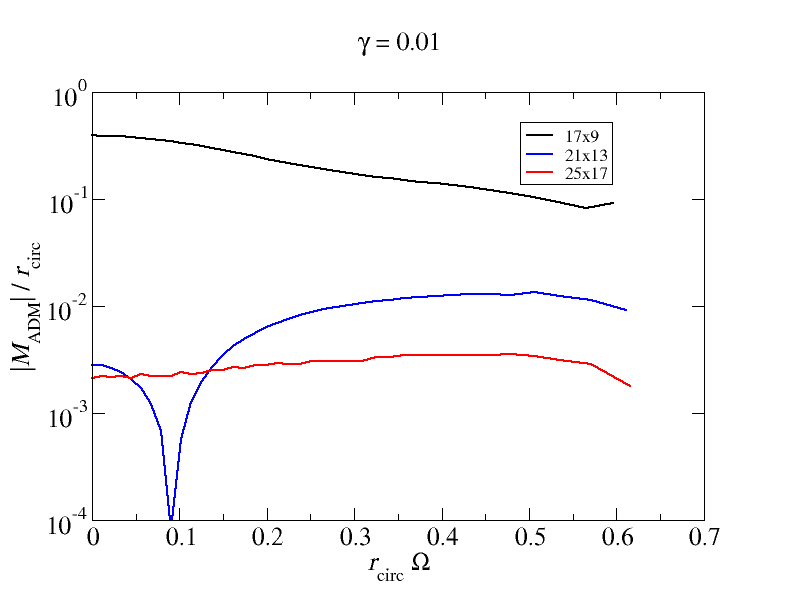}
        }
        {
            \includegraphics[width=0.47\textwidth]{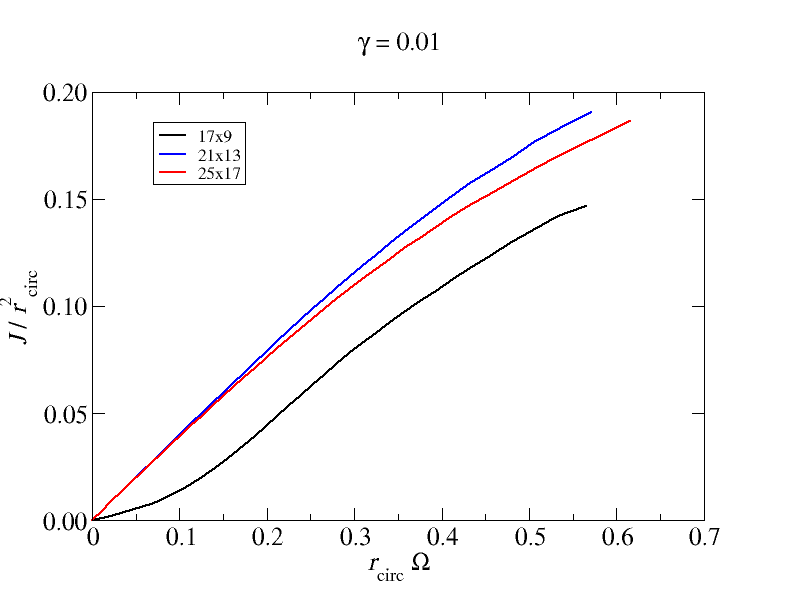}
        }
    \end{center}
\caption{The ADM mass $M_{\rm ADM}$ (first panel) and angular momentum $J$ (second panel), for a sequence of rotating solutions and three different resolutions. The coupling constant is fixed to $\gamma = 0.01$. The ADM mass converges to zero whereas the angular momentum is non-zero.}
\label{f:global}
\end{figure}

Figure \ref{f:global} shows, as an example, two global quantities for a sequence of rotating solutions: the ADM (Arnowitt-Deser-Misner) mass and the angular momentum. Both quantities are computed by surface integrals at infinity (see chapter 8 of \cite{Gourgoulhon:2007} for instance). The first panel shows that the ADM mass vanishes, for all configurations, as its values converges to zero when increasing the resolution (the blue curve behavior comes from a change of sign in the computed ADM mass). This is an effect already observed in \cite{VanAelst:2019} where it was shown that the Komar mass of the configurations must be zero as soon as the coupling constant $\gamma$ is different from zero. The configurations being stationary, the Komar and ADM mass must coincide. It implies that the ADM mass must also vanish, as is confirmed by the first panel of Fig. \ref{f:global}. The situation is different for the angular momentum shown in the second panel of Fig. \ref{f:global}. The curves do not converge to zero and the difference between the two highest resolutions gives a measure of the overall precision on the value of $J$.

\begin{figure}
    \begin{center}
        {
            \includegraphics[width=0.47\textwidth]{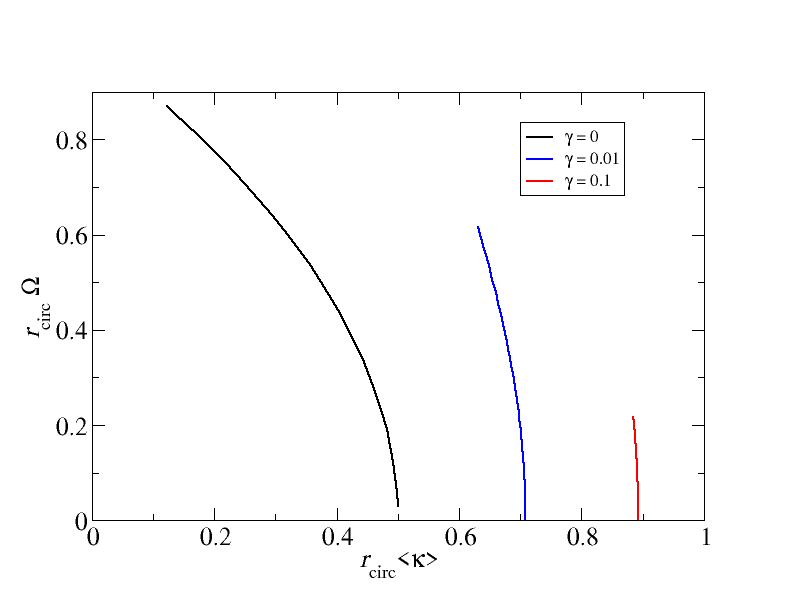}
        }
        {
            \includegraphics[width=0.47\textwidth]{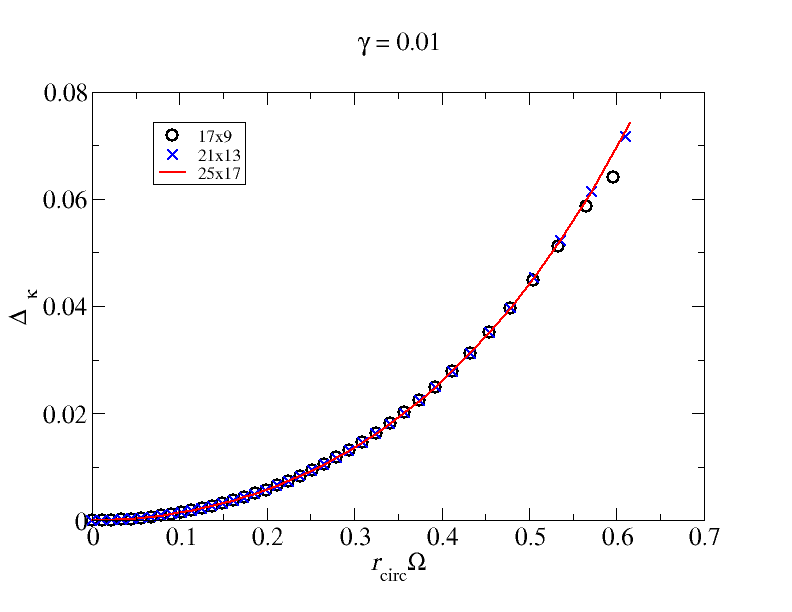}
        }
    \end{center}
\caption{The first panel shows the angular velocity $\Omega$, as a function of the average value of the surface gravity on the horizon. The second panel shows the average relative variation on the horizon of the surface gravity $\Delta_\kappa$, as a function of $\Omega$. Quantities are scaled by $r_{\rm circ}$.}
\label{f:kappa}
\end{figure}

To further illustrate the differences between the cubic Galileon black holes and the classical ones, one can study the surface gravity. At each point of the horizon, it can be computed by (see Eq. (10.9) in \cite{Gourgoulhon:2005}): 

\be
\label{e:kappa}
\kappa = \tilde{s}^i D_i N - N K_{ij} \tilde{s}^i \tilde{s}^j.
\ee

The first panel of Fig. \ref{f:kappa} shows the orbital velocity $\Omega$ as a function of the average value of $\kappa$ on the horizon. The black curve corresponds to the Kerr black hole case and to other ones to two different values of the coupling constant. The average is taken on the collocation points that lie on the horizon. So it is not the true angular average but converges rapidly to it, when resolution increases. The plot clearly indicates that the cubic Galileon field has a strong impact, at least on the average value of the surface gravity, even in the non-rotating case. Moreover, the effect of the cubic Galileon does not limit itself to the average of the surface gravity. It also makes the solutions deviate from the zeroth law of black holes thermodynamic which states that the surface gravity must be constant on the horizon. This can be measured by computing the mean of the absolute value of relative deviation on the horizon, defined as $\Delta_\kappa = \langle\displaystyle\frac{\l|\kappa - \langle\kappa\rangle\r|}{\kappa}\rangle$, quantity that vanishes if and only if $\kappa$ is constant. This deviation is shown on the second panel of Fig. \ref{f:kappa} for rotating sequences of various resolutions. The value of $\Delta_\kappa$ is clearly independent of the resolution and in particular it does not converge to zero: the black holes in the cubic Galileon theory don't obey the zeroth law.

It was shown in \cite{Bardeen:1973} that the zeroth law is true for stationary black holes under the dominant energy condition. For the configurations obtained in this paper, it turns out that the value of the energy density given by Eq. (\ref{e:E}) is negative everywhere. It violates the weak energy condition which states that $T_{\mu\nu} X^\mu X^\nu \geq 0$ for every time-like vector $X^\mu$, meaning every physical observer measures a positive energy density. The weak energy solution being included in the dominant one, the latter is also violated~;~there is no reason the zeroth law should hold. This can be linked to the positive energy theorem \cite{Schoen:1979, Schoen:1981, Witten:1981} which essentially states that any spacetime with zero ADM mass and that obeys the dominant energy condition must be Minkowski spacetime. It follows that the zero ADM mass configurations constructed in this paper cannot obey the dominant energy condition and so can violate the zeroth law, as is observed. 

 \section{Conclusion}

In this paper, for the first time, exact configurations (up to the numerical precision) of rotating black holes in the cubic Galileon theory are constructed. This is achieved by moving away from the quasi-isotropic coordinates used before. Instead a set of differential gauges under the 3+1 formalism is used : maximal slicing for the time coordinate and the spatial harmonic gauge for the spatial coordinates. In this context, the presence of the black hole is enforced by demanding that it is an apparent horizon in equilibrium. The full set of boundary conditions that ensues from those choices has been presented is details in \cite{Grandclement:2022} and used with success in several cases. The fact that the equation for the scalar-field contains third order derivatives (instead of two for more usual cases), leads to additional complication in terms of what boundary conditions must be used for the scalar-field. Nevertheless, an appropriate choice has been found and enabled the successful computation of rotating black holes in the cubic Galileon theory.

After carefully assessing the validity of the numerical results by monitoring various error indicators, especially those linked to the gauge choice, several properties of the solutions have been discussed. In particular, as in \cite{VanAelst:2019}, it is shown that the mass of the black hole vanishes, contrary to the angular momentum. A link is made between that property and the fact that the configurations do not obey the zeroth-low of thermodynamics. Both features arise from the fact that the stress-energy tensor coming from the scalar-field doesn't verify the dominant energy condition.

If the black holes in the cubic Galileon theory do not seem to be a valid alternative to the astrophysical ones, especially with a vanishing mass, they are still worth studying. First, there was a need to cure the main limitation of \cite{VanAelst:2019}, limitation coming form the use of quasi-isotropic coordinates. Second, this paper is another successful application of the formalism presented in \cite{Grandclement:2022}, a valuable tool to compute black hole solutions in various context. For the future, there are plans to apply it to cases that are still allowed by the observations. Once computed, various physical observables could be extracted from the numerical solutions. One could study the orbits of massive or massless particles around those objects, accretion disks in their vicinity, compute the frequencies of the quasi-normal modes or extract the gravitational waveform emitted by a binary system. In the long run, it will help unveiling the nature of the most compact objects in the Universe and put constraints on the theory of gravity.

\ack{The author acknowledges the support of the French Agence Nationale de la Recherche (ANR), under grant ANR-22-CE31-0015 (project StronG). This work was granted access to the HPC
resources of MesoPSL financed by the Region Ile de
France and the project Equip@Meso (Reference
No. ANR-10-EQPX-29-01) of the programme
Investissements d’Avenir supervised by the Agence
Nationale pour la Recherche.}

\section*{References}
\bibliographystyle{iopart-num}
\bibliography{mybib.bib}

\end{document}